\documentclass[acmsmall,screen,nonacm]{acmart}

\AtBeginDocument{%
  }

\usepackage{qcircuit}
\usepackage{wrapfig}
\usepackage{xcolor,listings}
\usepackage{algorithm,algpseudocode}
\usepackage{multirow}
\usepackage{graphicx}
\usepackage{booktabs}
\usepackage{tikz}
\usepackage{pgfplots}
\usepackage{listings}
\pgfplotsset{compat=1.18}

\def\({\begin{eqnarray*}}
\def\){\end{eqnarray*}}

\def\ket#1{|#1\rangle}

\algrenewcommand\algorithmicrequire{\textbf{Input:}}
\algrenewcommand\algorithmicensure{\textbf{Output:}}

\newcommand{\y}{$\checkmark$}
\newcommand{\n}{$\circ$}
\newcommand{\p}{(\y)} 

\usepackage{soul}

\usetikzlibrary{backgrounds,calc}

\definecolor{bg}{RGB}{248,248,248}
\definecolor{keyword}{RGB}{0,0,180}
\definecolor{comment}{RGB}{0,128,0}
\definecolor{string}{RGB}{186,33,33}
\definecolor{number}{RGB}{102,102,102}
\definecolor{identifier}{RGB}{0,0,0}

\lstdefinestyle{minted}{
    basicstyle=\ttfamily\normalsize,
    keywordstyle=\color{keyword}\bfseries,
    commentstyle=\color{comment}\itshape,
    stringstyle=\color{string},
    identifierstyle=\color{identifier},
    showstringspaces=false,
    breaklines=true,
    breakatwhitespace=true,
    tabsize=4,
    keepspaces=true,
    frame=single,
    framerule=0pt,
    rulecolor=\color{bg},
    xleftmargin=1em,
    framexleftmargin=1em,
    columns=fullflexible,
    upquote=true
}

\lstdefinelanguage{qsharp}{
  morekeywords={
    open, operation
  },
  sensitive=true
}

\lstdefinelanguage{silq}{
  morekeywords={
    vector, measure, while, def, lifted, for, in, if, return
  },
  morecomment=[l]{//},
  sensitive=true
}

\begin{document}

\title{A Survey of Quantum Programming Languages}


\author{Quan Do}
\email{biquando@cs.ucla.edu}
\orcid{0009-0006-5211-8678}
\author{Hersh Gupta}
\email{hershgupta@g.ucla.edu}
\author{Xiyuan Cao}
\email{xiyuan23@g.ucla.edu}
\orcid{0009-0005-9221-1597}
\author{Aarav Pabla}
\email{apabla@g.ucla.edu}
\orcid{0009-0004-3680-4788}
\author{Pranav Singamsetty}
\email{pranavsingam@g.ucla.edu}
\author{Evan O'Grady}
\email{eogrady01@g.ucla.edu}
\author{Keli Huang}
\orcid{0009-0003-9104-0621}
\email{kelihuang@cs.ucla.edu}
\author{Jens Palsberg}
\email{palsberg@ucla.edu}
\orcid{0000-0003-4747-365X}
\affiliation{%
  \institution{University of California, Los Angeles}
  \city{Los Angeles}
  \state{CA}
  \country{USA}
}

\renewcommand{\shortauthors}{Do, Gupta, Cao, Pabla, Singamsetty, O'Grady, Huang, and Palsberg}

\begin{abstract}
Quantum computing has seen multiple recent breakthroughs and is getting closer to demonstrations of an exponential advantage over classical computing for certain problems. Programmers will require high-level, general-purpose, executable programming languages to express quantum solutions clearly and effectively, and the field has already produced a wide variety of such languages. This paper presents a language classification framework and uses it to survey ten popular quantum programming languages. The findings include conceptual and experimental comparisons that result in a list of challenges for future language design.
\end{abstract}

\ccsdesc[500]{Computer systems organization~Quantum computing}
\ccsdesc[500]{Software and its engineering~Software notations and tools}

%

\keywords{programming languages, benchmark programs}


\maketitle

\section{Introduction}
\label{sec:introduction}

\paragraph{Context.}
Quantum computing has progressed from thought experiments and small proof-of-concept
demonstrations to systems with more than a thousand qubits.
As hardware scales, programmers will increasingly rely on high-level programming
languages to express algorithms, reason about correctness, and integrate quantum
subroutines into larger classical software systems. 
This will complement the foundations
laid out in the standard texts and the early work on quantum algorithms and simulation~\cite{nielsen_chuang_2010,lloyd_universal_1996}.

\paragraph{Scope.}
During the last decade, a diverse ecosystem of quantum programming languages has emerged. 
This paper focuses on high-level, general-purpose, executable languages that are available as actively maintained open-source software (as of 2026). We cover ten languages, originating both from industry (Cirq [Google], CUDA-Q [NVIDIA], Guppy [Quantinuum], PennyLane [Xanadu], PyQuil [Rigetti], Q\# [Microsoft], Qiskit [IBM], and Qualtran [Google]) and from academia (Qrisp [Fraunhofer]
and Silq [ETH Zurich]).


\paragraph{Workloads.}
Among the many algorithmic workloads for which quantum languages are intended,
Shor's algorithm for factoring integers and Hamiltonian simulation of quantum systems are of paramount importance.
They are both theoretically central and practically important, underpinning applications in cryptography and science~\cite{feynman_simulating_1982,berry_simulating_2015}.
At the same time, they stress many aspects of language design,
including the concise expression of complex algorithms, interaction with classical computation, and ease of verification.

Existing surveys of quantum programming languages rarely compare concrete programs for common tasks, and
most of them predate the languages that are in active use today~\cite{gay_qpl_survey_2006,sofge_qpl_survey_2008,heidt_qpl_survey_2020}.
In contrast, this paper uses Shor's algorithm and Hamiltonian simulation as running examples.
Specifically, in ten popular languages, we provide implementations of integer factoring and two different ways of simulating two different Hamiltonians, for a total of $1 + 2 \times 2 = 5$ programs in each language.

\paragraph{Contributions.}
This paper makes three main contributions:
\begin{enumerate}
  \item We introduce a classification framework for quantum programming languages with three dimensions: programming model, expressiveness, and safety, after which we apply the framework to ten languages.
  \item We provide implementations of 50 benchmark programs and an automated cross-language testing harness that measures program size in lines of code, program compilation and execution time, and program correctness against a classical baseline. All of our code is available on GitHub \cite{github_repo}.
  \item We present both conceptual and experimental comparisons across languages, highlighting common design patterns, points of friction, and open challenges for future languages.
\end{enumerate}

\paragraph{Organization.}
Section~\ref{sec:background-and-foundations} reviews background on quantum
programming, Shor's algorithm, and Hamiltonian simulation, and
Section~\ref{sec:classification-framework} introduces our classification
framework.
Section~\ref{sec:ten-quantum-languages} introduces the ten languages and presents a table with program sizes.
Section~\ref{sec:conceptual-comparison} synthesizes conceptual lessons, focusing on programming model, expressiveness, and safety.
Section~\ref{sec:experimental-comparison} describes our experimental
methodology and results.
Section~\ref{sec:open-challenges-and-future-directions} discusses open
challenges and future directions, and
Section~\ref{sec:conclusion} concludes.

\section{Background}
\label{sec:background-and-foundations}

This section introduces quantum computation, Shor's algorithm, and Hamiltonian simulation.

\paragraph{Quantum Concepts.}

The state of an $n$-qubit system can be represented by a unit vector $\ket{\psi}$ of $2^n$ complex numbers.  
A computation step from a state to a state can be specified by 
a unitary matrix $U$ of size $2^n \times 2^n$, and 
the readout from a state is done via projective measurement.

\paragraph{Quantum Programming Languages.}

Early quantum algorithms such as Shor's factoring algorithm and Grover's
search algorithm were typically presented in terms of circuit diagrams or pseudo-code.
In a circuit diagram, a program is specified by a sequence of unitary operations that, in many cases, work on one or two qubits.
The first quantum programming languages tended to focus on core ideas, emphasizing semantics over practical executability on a quantum computer.
More recently, industrial tool chains and open-source ecosystems have changed the direction of the field towards executability.
We pick up on this by surveying executable quantum programming languages, all of which can be compiled to circuits.



\paragraph{Shor's Algorithm.}
The goal of Shor's algorithm is to factor an integer $N$ using a hybrid of classical and quantum computation. Specifically, the classical part determines an integer $a$ such that $\gcd(a,N) = 1$ and then
calls a quantum subroutine:
\[
{\mbox FindOrderCandidate}(a,N)
\]
which attempts to return the order of $a$ in ${\mathbb{Z}}_N$, defined as the smallest integer $r>0$ such that $a^r \equiv 1\ (\mbox{mod}\ N)$.  However, both the call to ${\mbox FindOrderCandidate}(a,N)$ and the subsequent classical processing to determine a factor of $N$ may fail, in which case Shor's algorithm repeats until either success or timeout.
Shor's algorithm is exponentially faster than any known classical algorithm because ${\mbox FindOrderCandidate}(a,N)$ does its job exponentially faster than any known classical method.

\begin{figure}
  \centering
  \includegraphics[width=0.8\linewidth]{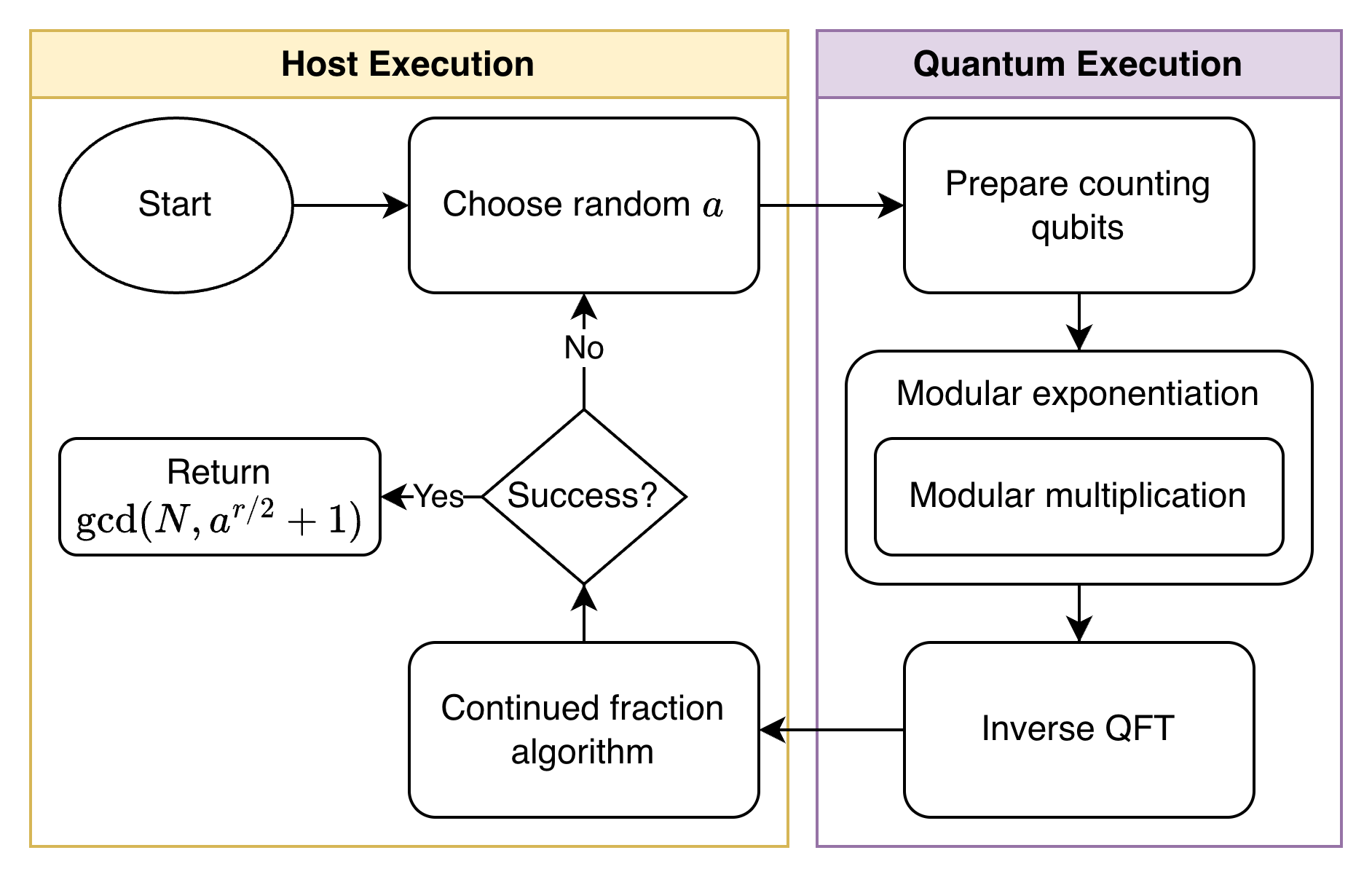}
  \Description{}
  \caption{A sketch of Shor's algorithm. The order finding procedure consists of the quantum circuit as well as the classical continued fraction algorithm.}
  \label{fig:algo-shors}
\end{figure}

Figure~\ref{fig:algo-shors} shows a high-level outline of Shor's algorithm. For simplicity, we assume that $N$ is a product of two prime numbers. The program starts by choosing the random integer $a$, and checking that it meets certain conditions. Then it runs a quantum routine as well as the classical continued fraction algorithm to find the order $r$ of $a$ in $\mathbb{Z}_N$. Finally, we check if $\mathrm{gcd}(N,a^{r/2}+1)>1$. If so, we have obtained a factor of $N$. Otherwise, we restart the algorithm with a new choice of $a$.

The quantum routine involves preparing a $t$-qubit 
counting register $\ket{c}$, performing modular exponentiation on ancilla qubits ($\ket{1}\mapsto\ket{a^c\text{\ mod\ }N}$), and performing an inverse Quantum Fourier Transform, which yields $2^tj/r$ for some integer $j<r$. Finally, the classical continued fraction algorithm can compute $r$ from this value.

In Section~\ref{sec:conceptual-comparison}, we give examples of how to implement each step of the quantum routine in Figure~\ref{fig:algo-shors}.

\paragraph{Hamiltonian Simulation.}

A Hamiltonian is a Hermitian matrix that governs the evolution of a quantum system according to the Schr\"odinger equation.
Given a Hamiltonian $H$ acting on $n$ qubits, the goal of
Hamiltonian simulation is to approximate the unitary time-evolution 
$U(t) = e^{- \mathrm{i} H t}$ applied to an initial state.
In this paper, we focus on two widely-studied Hamiltonians:
\begin{itemize}
  \item the transverse-field Ising model (TFIM),
        $H_{\mathrm{TFIM}} = J \sum_{i} Z_i Z_{i+1} + h \sum_i X_i$, and
  \item the Heisenberg isotropic chain with a longitudinal field,
        $H_{\mathrm{Heis}} =
           J \sum_i (X_i X_{i+1} + Y_i Y_{i+1} + Z_i Z_{i+1})
           + B \sum_i Z_i$.
\end{itemize}
Here, $X_i,Y_i,Z_i$ denote the single-qubit Pauli operators acting on site $i$ of the
spin chain.  Additionally, in TFIM, $J$ is the Ising coupling and $h$ is the transverse field strength, while in the Heisenberg model, $J$ in the exchange coupling and $B$ is the longitudinal field strength.


These Hamiltonians appear prominently in recent benchmark suites and libraries
for Hamiltonian simulation.
For example,
HamLib \cite{hamlib_sawaya_2024} collects standardized instances of TFIM, Heisenberg-type spin chains,
and related models as part of a curated library of Hamiltonians for
benchmarking algorithms and hardware, while
AppQSim \cite{appqsim_granet_2025} and follow-up cross-model studies \cite{crossmodel_chatterjee_2025} use TFIM and Heisenberg chains as representative workloads when comparing simulation methods and
platforms.
Aligning our benchmarks with these public datasets gives us canonical test cases that can potentially be extended to more challenging Hamiltonians, such as Fermi-Hubbard plaquettes.
Both our benchmarks are simple enough to be programmed in many languages with reasonable effort, yet rich enough to exercise nontrivial language constructs and lead to insight.

We consider two algorithmic approaches to Hamiltonian simulation:
Trotterization \cite{trotter_product_1959} and Linear Combination of Unitaries (LCU) \cite{berry_simulating_2015}. Both approaches are exponentially faster at simulating an arbitrary Hamiltonian than any known classical algorithm because they represent the state vector by $n$ qubits, rather than by a vector of length $2^n$, and operate on those state vectors in time that is polynomial in $n$.
We apply both approaches to simulation of both Hamiltonians.

\begin{figure}
  \centering
  \includegraphics[width=0.9\linewidth]{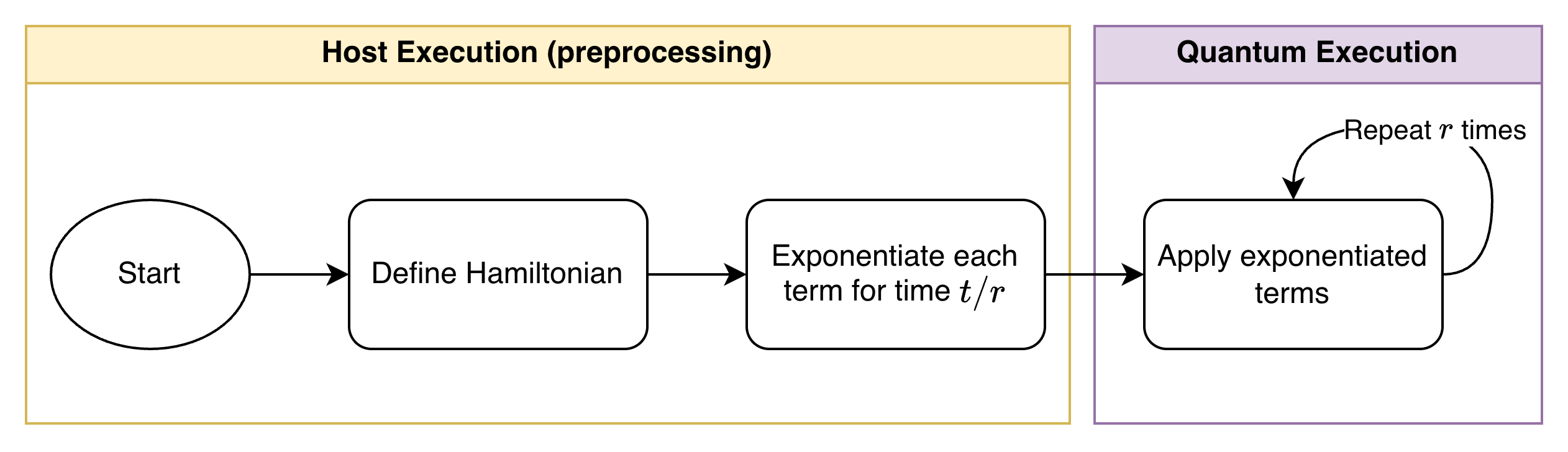}
  \Description{}
  \caption{A sketch of the Trotterization approach to Hamiltonian simulation.}
  \label{fig:algo-trotter}
\end{figure}

\paragraph{Trotterization}
If we write the Hamiltonian as a sum of terms 
$H = \sum_k H_k$, a first-order Lie-Trotter product formula with $r$ steps \cite{trotter_product_1959} 
approximates the time evolution as:
\[
  e^{-\mathrm{i} H t}
  \approx
  \left( \prod_k e^{-\mathrm{i} H_k t / r} \right)^{r}
\]
Trotterization implements the right-hand side on a quantum computer. The $H_k$ terms are chosen so that each $e^{-iH_kt/r}$ can be efficiently calculated on a classical computer and implemented in a quantum computer (see Figure~\ref{fig:algo-trotter}). There exist higher order Trotter formulas that increase the approximation accuracy at the cost of a longer program, but in this paper we implement the first order formula shown above. The error in this formula scales as $O(t^2/r)$.

Figure~\ref{fig:algo-trotter} shows an outline of the Trotterization algorithm. First, the Hamiltonian $H$ is defined as a sum of local Pauli operators $H_k$. Then each term is exponentiated to get $e^{-\mathrm{i}H_kt/r}$. Finally, we apply the exponentiated terms in the quantum routine, repeating $r$ times. 

In Section~\ref{sec:conceptual-comparison}, we give examples of how to implement both the classical steps and the quantum step in Figure~\ref{fig:algo-trotter}.

\paragraph{Linear Combination of Unitaries.}
For LCU, we 
truncate the Taylor expansion of $e^{-\mathrm{i} H t}$~\cite{berry_simulating_2015}, which gives us this approximation:
\begin{equation}
  e^{-\mathrm{i} H t}
  \approx
  I - \mathrm{i} t H - \tfrac{1}{2} t^2 H^2
\label{eq:truncated-taylor-expansion}
\end{equation}
The LCU approach implements the right-hand side on a quantum computer.  Notice that the elements on the right-hand side are not necessarily unitary and include $H^2$, which is the square of a matrix of size $2^n \times 2^n$.  Those observations play a role in an implementation of the LCU approach. LCU has a lower error bound than Trotterization, at the cost of increased complexity and required ancilla qubits.

\begin{figure}
  \centering
  \includegraphics[width=0.9\linewidth]{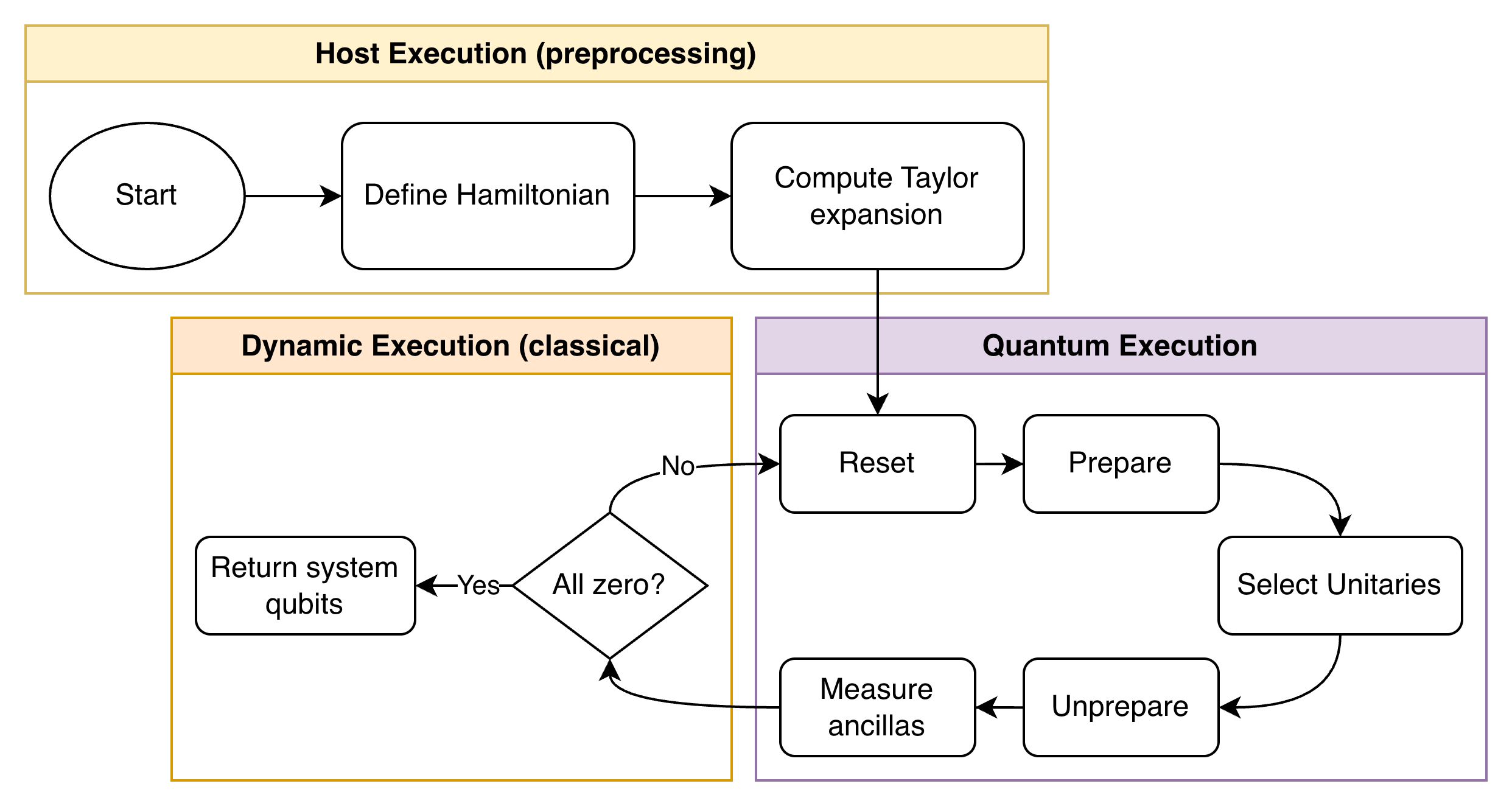}
  \Description{}
  \caption{A sketch of the LCU approach to Hamiltonian simulation. The repeat-until-success (RUS) procedure consists of the quantum circuit and the classical mid-circuit logic.}
  \label{fig:algo-lcu}
\end{figure}

Figure~\ref{fig:algo-lcu} shows a diagram of the LCU algorithm. First, the Hamiltonian $H$ is defined as a sum of Pauli operators. 
Next, the truncated Taylor expansion of $e^{-iHt}$ in Equation~\ref{eq:truncated-taylor-expansion} is computed classically. Since the $H$ and $H^2$ terms can be represented as a sum of Pauli operators, the whole expression is a linear combination of unitary matrices.
In the quantum routine, we prepare an ancilla register into a state that encodes the complex coefficients of the linear combination, selectively apply the unitary operators, reverse the ancilla preparation, and measure the ancilla qubits. If all ancilla qubits are zero, then the algorithm is successful, and otherwise, the quantum routine is repeated. 
This algorithm constructs a quantum circuit that can be represented as a block matrix, where the linear combination of unitaries is in the top-left corner. This procedure, called block encoding, encodes a classically-obtained matrix into the quantum circuit.

This algorithm requires mid-circuit classical logic to determine whether the quantum subroutine should repeat based on some measurement outcomes. The quantum circuit along with the classically-conditioned repetition is commonly referred to as a repeat-until-success (RUS) procedure.

In Section~\ref{sec:conceptual-comparison}, we give examples of how to implement all the steps in Figure~\ref{fig:algo-lcu}.

\paragraph{High-Level, General-Purpose, Executable Quantum Programming Languages.}

In the world of executable quantum programming languages, we focus on high-level, general-purpose languages.
We define general-purpose loosely to mean that we can easily program both Shor's algorithm and two algorithms for Hamiltonian simulation.    We set the bar for being high-level fairly low and define it to mean that we can easily program our chosen algorithms in ways that take any input of any size.  
Specifically, we can program Shor's algorithm such that it can factor any integer, and we can program an algorithm that simulates a Hamiltonian of any size. 
However, the idea of a high-level language involves much more than taking input of any size, and this is where our classification framework comes in.

We left out some interesting languages that don't satisfy all our criteria.
In particular, we left out 
Qmod \cite{GarroRamirezCorrales25} which is not available as open-source software, 
Quipper \cite{quipper_paper} which is no longer actively maintained, and 
Strawberry Fields \cite{strawberryfields_docs,
strawberryfields_paper} which was deprecated in early 2026.
We also left out OpenQASM~3 \cite{openqasm3_paper} which is a low-level language with poor support for parameterization, and HML \cite{simuq_paper} which is a domain-specific language for Hamiltonian simulation that leaves little room for implementing Shor's algorithm.
Finally, we left out pytket \cite{tket_docs,tket_paper}, which is Quantinuum's older interface for writing quantum programs before they introduced Guppy.

\paragraph{Terminology.}

A program execution can be divided into different parts depending on where and when the operations are executed. Table~\ref{tab:terminology} outlines the terminology we use for each part of a program execution. In particular, there are three disjoint sections of a program: host execution, dynamic execution, and quantum execution. Quantum execution consists of purely quantum operations, while host and dynamic execution consist of classical operations.

\begin{figure}
  \centering
  \includegraphics[width=0.6\linewidth,trim={0 2.0cm 0 0},clip]{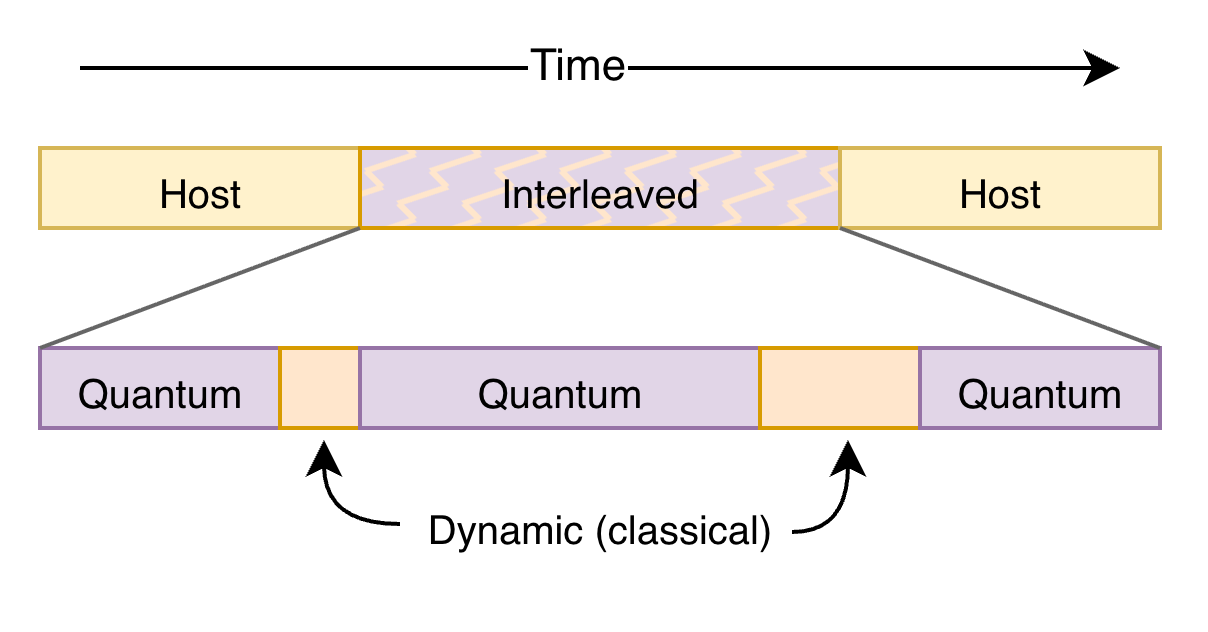}
  \Description{}
  \caption{A timeline of the overall program execution.}
  \label{fig:timeline}
\end{figure}

Figure~\ref{fig:timeline} depicts a timeline of the overall program execution, showing when each section executes. First, the host performs some classical preprocessing. Then it initializes an interleaved quantum-classical routine. The interleaved routine has some quantum operations as well as some dynamic classical operations. Then the host receives the results, usually in the form of quantum measurements, and performs some classical postprocessing. The host can possibly initialize additional interleaved routines (like in Figure~\ref{fig:algo-shors}).

\begin{table*}[t]
\centering
\caption{Terminology for Each Part of a Program Execution.}
\begin{tabular}{l|p{9cm}}
  \toprule
  Name & Description \\
  \midrule
  Overall program execution & The entire program execution, including host execution, dynamic classical execution, and quantum execution. \\
  Host execution & The classical part of the overall program execution that runs either before a quantum routine has started execution (preprocessing) or after it has finished (postprocessing). Quantum data does not exist during this time. Runs on any classical computer. \\
  Dynamic (classical) execution & The classical part of the overall program execution that executes in response to and provides control for quantum operations. Runs on a classical co-processor concurrently with the quantum processor. \\
  Quantum execution & The purely quantum part of the overall program execution. Runs entirely on a quantum processor. \\
  Interleaved execution & Quantum execution along with dynamic classical execution. This is the part of the overall program execution where quantum and classical operations may be interleaved. \\
  \bottomrule
\end{tabular}
\label{tab:terminology}
\end{table*}

\section{Classification Framework}
\label{sec:classification-framework}

We now introduce the framework that we use to classify and compare high-level, general-purpose, executable quantum programming languages.  
The framework refines the notion of high-level into three dimensions: programming model, expressiveness, and safety.

\subsection{Programming Model}

All of our example algorithms have both classical and quantum aspects, as we described in Section~\ref{sec:background-and-foundations}.  
Such a mix of classical and quantum aspects has been supported by quantum programming languages in different ways, and here we present two ends of a spectrum.  
\begin{itemize}
    \item Embedded: a classical program builds a quantum program as a data structure and invokes a routine to execute this quantum program either on a quantum computer or on a simulator.
    \item Hybrid: a program seamlessly blends both computation steps that eventually will run on a classical computer and computation steps that eventually will run on a quantum computer.  
\end{itemize}
For example, for Shor's algorithm, if we use an embedded language, we will write code for $\gcd$ and other classical processing as usual, and we will build the quantum routine $\mbox{FindOrderCandidate}(a,N)$ as a data structure.
Indeed, such a language has two levels, where the classical level builds and invokes code for the quantum level.
The programmer prepares interleaved operations by constructing a data structure, either implicitly or explicitly, onto which quantum and dynamic classical operations are appended. 

In contrast, if we use a hybrid language, we will write the code for $\gcd$, $\mbox{FindOrderCandidate}(a,N)$, and more as statements at the same level.
Such a language has a single level, where all code is executed without regard to whether we think of computation steps as classical or quantum.  
%
The programmer seamlessly interleaves quantum and classical operations, and may perform classical operations as a result of one part of the quantum data, while keeping the other quantum data intact. 

We summarize the differences between the two programming models in three points.

\begin{itemize}
\item
\textbf{Separation:}
Embedded languages enforce that the programmer knows exactly which operations occur during host execution, dynamic execution, and quantum execution. Hybrid languages may interleave classical and quantum operations within the same subroutine, though they may still require the programmer to distinguish host and dynamic operations.

\item
\textbf{Program Representation:}
Embedded languages encode quantum programs as a data structure, onto which operations are appended. Hybrid languages avoid such a data structure.


\item
\textbf{Compilation:}
Embedded languages require the programmer to build a quantum program structure that can be compiled to the circuit level. Hybrid languages require the compiler to determine how to run each operation and ensure that the quantum operations are realizable.
\end{itemize}


\subsection{Expressiveness}
We use the term \emph{expressiveness} to refer to the abstractions and
language constructs that enable easy programming of Shor's algorithm and Hamiltonian simulation.  We will cover four expressiveness ideas that are  helpful for making our benchmark programs concise and readable.
\begin{itemize}
    \item Support for Pauli manipulation: Pauli operator types, Pauli algebra, and Pauli exponentiation.
    \item Encoding of classical data: state preparation and block encoding.  
    \item Quantum integers: computation with integers represented as collections of qubits.
    \item Dynamic allocation of qubits.
\end{itemize}
For example, Pauli manipulation is useful in both algorithms for Hamiltonian simulation. In Trotterization, the desired Hamiltonian is broken up into a sum of Pauli terms, each of which is exponentiated, yielding a sequence of unitary operations that can be implemented as quantum gates. This algorithm can be expressed in a simple, declarative way using a language that supports construction of a Hamiltonian and exponentiation of its terms. Similarly, LCU is easy to implement in a language that has good support for constructing the Taylor expansion of the Hamiltonian.


The LCU approach uses a state-preparation circuit that encodes the complex-valued coefficients of the Taylor expansion of the Hamiltonian into the ancilla qubits' state vector. 
It also block-encodes some of the Hamiltonian's Taylor expansion into the top-left corner of a unitary matrix. 

The main part of Shor's algorithm involves performing modular exponentiation in the quantum circuit. There are several ways to program this procedure, but one common method is via arithmetic operations on quantum integers \cite{pavlidis2012fast}. This part of Shor's algorithm is easy to implement in a quantum programming language that natively supports quantum integers and operations on them.

Many aspects of our workloads are implemented with the use of helper qubits. Similarly to classical programming, the best practice is to use helper qubits with a short lifetime and in a localized portion of the program.  This can be achieved by allocating helper qubits dynamically just before they are needed, and de-allocating them right after they are no longer needed. Dynamic allocation of helper qubits is not strictly necessary in our algorithms because we can calculate the number of required qubits ahead of time. However, some of our LCU and Shor's algorithm implementations use dynamic allocation to simplify the programs or make them more declarative.

\subsection{Safety}
We take safety to mean language mechanisms and tool support that make it harder to write wrong programs. We will cover two safety ideas that stem from classical languages. 
\begin{itemize}
    \item Static type safety of the quantum code: the program can apply an operation or a procedure only to arguments of types for which it was intended, and this is checked statically before we run the program.
    \item Initialization: every variable is initialized to a known value, typically $\ket{0}$. This requires that the language ensures that quantum resources can be cleanly deallocated. Readers familiar with quantum computing know this idea as ``automatic uncomputation.''
\end{itemize}
Type safety ensures that we apply functions and operations only to values of the intended type. Static type safety goes further and ensures this property before the program is executed.  For example, static type safety ensures that a quantum operation that works on two qubits will always be applied to two qubits and never on one qubit or some other number of qubits.

In classical programming languages, it is often advantageous to prevent the programmer from declaring variables without initializing them. This prevents a variable from containing a garbage value that non-deterministically affects program behavior. In quantum programming, there is an analogous form of initialization safety, where temporary qubits are always initialized to a value that is known and prevents the qubit from being entangled with other qubits.


\subsection{Summary}
Embedded languages provide a clear separation between quantum and classical logic, whereas hybrid languages have more homogeneous programs. Pauli manipulation makes implementations of Trotterization and LCU more declarative, saving developer effort in manually calculating Pauli string exponentiations and Taylor expansions. Languages that provide state preparation and block encoding enable simpler implementations of LCU and other algorithms that encode classical vectors and matrices. Quantum integers make the modular exponentiation in Shor's algorithm easy to write and understand. Dynamic allocation can be useful for programs with dynamic control flow where quantum resource allocation cannot be easily done upfront. Finally, static type safety is useful for catching errors already during compilation, and initialization safety is useful for clean management of quantum resources.


\section{Ten Quantum Programming Languages}
\label{sec:ten-quantum-languages}

This section introduces the ten languages, grouped by whether they originate from industry or academia.  Table~\ref{tab:language-characteristics} highlights that all of them can run on a quantum computer and are available as actively maintained open-source software (as of 2026).
For a language to be executable on a quantum computer, it must either provide libraries for interfacing with such a computer or enable translation to a standard format such as OpenQASM. The number of commits is meant to show that every language is being actively maintained, rather than being a direct comparison between languages. Some languages have commits across multiple repositories, which is not reflected here.


\begin{table*}[h!]
  \centering
  \caption{The Ten Languages and Some of Their Characteristics}
  \label{tab:language-characteristics}
  \begin{tabular}{crlc|c|c|r}
    \toprule
    & & & 
    & \multicolumn{1}{c|}{runs on}
    & \multicolumn{1}{c|}{open}
    & \multicolumn{1}{c}{commits}
    \\
    Origin & Year & Language & Style 
    & \multicolumn{1}{c|}{a QC?}
    & source?
    & \multicolumn{1}{c}{in 2025$^1$}
     \\
    \midrule
    Industry 
    & 2018 & Cirq             & Python     
    & \y & \y & 619 \\
    & 2022 & CUDA-Q         & C++/Python 
    & \y & \y & 658 \\
    & 2024 & Guppy            & Python     
    & \y & \y & 320 \\
    & 2018 & PennyLane        & Python     
    & \y & \y & 1,462 \\
    & 2017 & PyQuil           & Python     
    & \y & \y & 18 \\
    & 2017 & Q\#              & C\#        
    & \y & \y & 446 \\
    & 2017 & Qiskit           & Python     
    & \y & \y & 802 \\
    & 2023 & Qualtran         & Python     
    & \y & \y & 133 \\
    \hline
    Academia
    & 2022 & Qrisp            & Python     
    & \y & \y & 1,575 \\
    & 2020 & Silq             & Imperative 
    & \y & \y & 307 \\
    \bottomrule
  \end{tabular}
  \caption*{$^1\,$Number of commits on the default branch of their GitHub repository}
\end{table*}

\begin{table*}[t!]
  \centering
  \caption{The Ten Languages and the Lines of Source Code for Our Benchmark Programs}
  \label{tab:program-size}
  \begin{tabular}{l|r|rr|rr}
    \toprule
             & Factoring & \multicolumn{2}{c|}{TFIM} & \multicolumn{2}{c}{Heisenberg} \\
    Language & Shor's & Trotter & LCU & Trotter & LCU \\
    \midrule
    Cirq             & 127\,\, & 45 & 203 & 53 & 203 \\
    CUDA-Q           & 114$^1$ & 38 & 196 & 40 & 191 \\
    Guppy            & 144$^1$ & 45 & 215 & 62 & 216 \\
    PennyLane        & 48\,\,  & 36 & 180 & 37 & 181 \\
    PyQuil           & 182$^1$ & 50 & 216 & 58 & 217 \\
    Q\#              & 173\,\, & 24 & 316 & 42 & 318 \\
    Qiskit           & 136$^1$ & 33 & 195 & 35 & 196 \\
    Qualtran         & 111\,\, & 35 & 181 & 64 & 183 \\ \hline
    Qrisp            & 50\,\,  & 61 & 149 & 63 & 150 \\
    Silq             & 87\,\,  & 15 & 137 & 25 & 140 \\
    \bottomrule
  \end{tabular}
  \caption*{
    $^1\,$These implementations use a permutation matrix, rather than quantum modular arithmetic.
  }
\end{table*}

\subsection{Languages from Industry}

We begin with eight languages from industry that were introduced in 2017--2024.

\paragraph{Cirq.}
Google introduced Cirq in 2018 as a Python-based framework in which ordinary Python code constructs and transforms circuit objects \cite{cirq_docs}.
Cirq is designed to be especially well suited for hardware-aware circuit construction, compilation, and experimentation with device-native gate sets on near-term processors.

\paragraph{\mbox{CUDA-Q}.}
NVIDIA introduced CUDA-Q in 2022 as a C++-centric platform with Python bindings for performance-oriented programming combined with Python for orchestration \cite{kim2023cuda}.
CUDA-Q is designed to support large-scale hybrid workflows that integrate quantum kernels with high-performance classical computation and accelerated simulation.

\paragraph{Guppy.}
Quantinuum introduced Guppy in 2024 as a statically typed language for structured programming in the style of Python and with strong compile-time guarantees~\cite{Koch2024,Koch2025}.
Guppy is designed to support the development of scalable quantum applications with disciplined handling of quantum data and control flow.

\paragraph{PennyLane.}
Xanadu introduced PennyLane in 2018 as a Python framework in the style of differentiable machine-learning libraries such as PyTorch and TensorFlow \cite{pennylane_docs}.
PennyLane was originally designed to enable hybrid quantum-classical optimization workflows, particularly for variational algorithms and quantum machine learning, but has since evolved into a more general framework.

\paragraph{PyQuil.}
Rigetti introduced PyQuil in 2017 as a Python interface to the Quil instruction language, in the style of Python APIs to build and execute programs \cite{pyquil_docs,quil_paper}.
PyQuil is designed to facilitate the compilation and execution of gate-based quantum circuits on Rigetti hardware and simulators.

\paragraph{Q\#.}
Microsoft introduced Q\# in 2017 as a stand-alone quantum programming language, in the style of C\# with structured operations and strong typing, and with support from .NET tooling \cite{qsharp_docs,qsharp_paper}.
Q\# is designed to support the development of scalable quantum algorithms by unifying quantum and classical programming.

\paragraph{Qiskit.}
IBM introduced Qiskit in 2017 as a Python-based framework with a large collection of libraries \cite{qiskit_docs,qiskit_paper,qiskit_algorithms}.
Qiskit provides an end-to-end toolchain from high-level algorithm description through compilation to execution on IBM Quantum hardware. 

\paragraph{Qualtran.}
Google introduced Qualtran in 2023 as a Python library for algorithm design, in the style of compositional software engineering with reusable components \cite{qualtran_paper}.
Qualtran is designed to support modular construction and resource estimation of large, fault-tolerant quantum algorithms.
One of the implementations compiles a Qualtran program to a Cirq program.

\subsection{Languages from Academia}

We continue with two languages from academia that were introduced in 2020--2022.

\paragraph{Qrisp.}
Fraunhofer introduced Qrisp in 2022 as a Python-embedded framework that enables programming with powerful abstraction mechanisms \cite{qrisp_docs}.
Qrisp is designed to enable the development of structured quantum algorithms and to ensure compatibility with existing compilation toolchains.
Qrisp is now being further developed by Fraunhofer, IQM, and others.


\paragraph{Silq.}
Researchers at ETH Zurich introduced Silq in 2020 as a high-level quantum programming language, in the style of imperative languages with strong typing \cite{silq_paper}.
Silq is designed to simplify quantum programming by providing automatic uncomputation.

\subsection{Implementation of Shor's Algorithm and Hamiltonian Simulation}

Table~\ref{tab:program-size} gives a bird-eye view of the implementations of our chosen workloads, listing counts of lines of source code.
We computed these numbers by removing comments and blank lines and consolidating multi-line statements, with the goal of counting only the number of logical lines of code. Table~\ref{tab:program-size} highlights that in all the languages, LCU requires more code than Trotterization.



\section{Conceptual Comparison}
\label{sec:conceptual-comparison}

Our chosen languages are high-level, general-purpose, and executable. In Section~\ref{sec:classification-framework}, we refined the notion of high-level into a classification framework organized along three dimensions: programming model, expressiveness, and safety. We now use our framework to provide a conceptual comparison of the ten languages. Table~\ref{tab:lang-dims} summarizes how languages are positioned along these dimensions and their associated subcategories.

{
\begin{table*}[h!]
  \centering
  \caption{The Ten Languages and Their Classification in Our Framework}
  \begin{tabular}{l|cc|cccc|ccc}
    \toprule
             & \multicolumn{2}{c|}{Programming model}
             & \multicolumn{4}{c|}{Expressiveness}
             & \multicolumn{2}{c}{Safety} \\
    Language & Embedded & Hybrid & Pauli & Encode & Q-Int & Dyn
    & Type & Init \\
    \midrule
    Cirq      & \y\,\, & \n\,\, &   \y & \y\,\, &     \n & \n\,\, &   \n & \n    \\
    CUDA-Q    & \n\,\, & \y$^2$ &   \y & \p$^3$ & \n & \p$^6$ &   \y & \n    \\
    Guppy     & \n\,\, & \y$^2$ &   \n & \p$^4$ & \n & \y\,\, &   \y & \n    \\
    PennyLane & \y$^1$ & \n\,\, &   \y & \y\,\, &     \n & \p$^6$ &   \n & \n    \\
    PyQuil    & \y\,\, & \n\,\, &   \y & \n\,\, &     \n & \n\,\, &   \n & \n    \\
    Q\#       & \n\,\, & \y\,\, &   \n & \y\,\, &     \n & \p$^6$ &   \y & \n    \\
    Qiskit    & \y\,\, & \n\,\, &   \y & \y\,\, &     \n & \n\,\, &   \n & \n    \\
    Qualtran  & \y\,\, & \n\,\, &   \n & \y\,\, &     \y & \n\,\, &   \n & \n    \\
    \hline
    Qrisp     & \y$^1$ & \n\,\, &   \y & \y\,\, &     \y & \p$^6$ &   \n & \y    \\
    Silq      & \n\,\, & \y\,\, &   \n & \p$^5$ & \y & \y\,\, &   \y & \y    \\
    \bottomrule
  \end{tabular}
  \caption*{A check mark \y\ means the language has support, a circle \n\ means it does not have support, and a check mark in parentheses (\y) means it has partial support. The expressiveness dimensions are Pauli manipulation, encoding classical data, quantum integers, and dynamic qubit allocation. The safety dimensions are static type safety of the quantum code and initialization safety.\\
  $^1\,$PennyLane and Qrisp are embedded by default, but both of them provide libraries for hybrid programming: PennyLane has Catalyst and Qrisp has Jasp. \\
  $^2\,$CUDA-Q and Guppy have some separation between the host language (Python) and the compiled quantum-classical functions, but they are hybrid inside of those functions. \\
  $^3\,$CUDA-Q supports initializing the state vector of a quantum register, but it does not synthesize this as a preparation circuit. This means it cannot be used in LCU, which also needs a preparation circuit. \\ 
  $^4\,$Guppy supports state preparation via importing the pytket \texttt{StatePreparationBox}, but does not support it natively. \\
  $^5\,$Silq provides encoding primitives in a separate repository~\cite{silq_linalg_repo} that can be downloaded and imported. \\
  $^6\,$CUDA-Q, PennyLane, Q\#, and Qrisp allow the programmer to allocate qubits in the middle of a program, but prohibit allocation conditioned on mid-circuit measurements.
  }
  \label{tab:lang-dims}
\end{table*}
}

\subsection{Programming Model}

A quantum programming language's programming model determines how it combines classical and quantum computation. In an embedded language, the programmer constructs the interleaved quantum/classical routine as a data structure composed of quantum and dynamic classical operations. In a hybrid language, the programmer can perform operations the same way whether in a quantum or classical context.

We will illustrate the differences between these two programming models with the repeat-until-success procedure in LCU; see Figure~\ref{fig:algo-lcu}. This procedure consists of a quantum routine as well as a dynamic classical part that responds to measurements from the quantum part. Embedded and hybrid languages have different ways of expressing the interleaved execution as part of the overall program execution.

\subsubsection{Embedded Languages}
\label{sec:embedded-languages}

An embedded language encodes an interleaved quantum/classical routine as a data structure embedded within a classical host programming language.
It enforces a separation between classical and quantum operations.
Embedded languages include Cirq, PennyLane, PyQuil, Qiskit, Qualtran, and Qrisp. These are all Python-based frameworks in which the user builds a quantum program either explicitly or implicitly as a Python data structure.

Although we classify PennyLane and Qrisp as embedded, these two languages have libraries that enable programming in a hybrid style. PennyLane provides the Catalyst package and Qrisp contains the Jasp submodule, both of which trace code from their parent language, allowing quantum and classical operations to be seamlessly interleaved.

\begin{figure}[h]
\centering
%
%
%
%

\begin{lstlisting}[style=minted,language=Python]
q = QuantumRegister(2)
c = ClassicalRegister(2)
qc = QuantumCircuit(q, c)

with qc.while_loop((c[0], 0)):         # [All zero?]
    qc.reset(q)                        # [Reset]

    qc.h(q[0])                         # [Prepare]
    qc.cx(q[0], q[1], ctrl_state='0')  # [Select unitary (X)]
    qc.cz(q[0], q[1], ctrl_state='1')  # [Select unitary (Z)]
    qc.h(q[0])                         # [Unprepare]

    qc.x(q[0])  # invert loop condition
    qc.measure(q[0], c[0])             # [Measure]
\end{lstlisting}
\Description{}
\caption{An implementation in Qiskit of the repeat-until-success procedure of LCU with the matrix $0.5X+0.5Z$. This implementation uses \texttt{QuantumCircuit.while\_loop()} in its context manager form. The comments with square brackets correspond to nodes in the LCU description in Figure~\ref{fig:algo-lcu}.}
\label{fig:example-qiskit-embedded-simple-1}
\end{figure}


In Qiskit, repeat-until-success can be implemented by constructing a subroutine that contains the operations to be repeated, and adding it to the parent routine along with the classical loop condition using \verb|QuantumCircuit.while_loop()|. This function can also be used in a \verb|with| statement, which allows the programmer to construct the loop using syntax that is somewhat similar to a normal Python loop. However, operations in the loop body must be quantum; any classical operations will run only once during host execution. Figure~\ref{fig:example-qiskit-embedded-simple-1} shows a Qiskit implementation of the repeat-until-success procedure in LCU for $0.5X+0.5Z$.
In this example, \verb|qc| represents an interleaved routine consisting of quantum and dynamic classical operations. Inside of the \verb|with| block, Qiskit automatically creates a context in which quantum operations are silently appended to a subcircuit, rather than the parent \verb|qc| circuit. We prepare the coefficients $(0.5,0.5)$ in the qubit \verb|q[0]| using a Hadamard gate, and we select the $X$ and $Z$ unitaries using $CX$ and $CZ$ gates. We invert and measure \verb|q[0]|, looping if the qubit does not measure to zero. After the block, the constructed subcircuit is silently appended to \verb|qc| along with a classical repeat operation. 

This example highlights two features of embedded languages. First, they have a clear separation between host classical, dynamic classical, and quantum operations. In Qiskit, host-executed control flow uses Python keywords like \texttt{if} and \texttt{while}, dynamic control is defined with circuit operations like \verb|QuantumCircuit.while_loop()|, and quantum control is specified using controlled quantum gates. Second, the interleaved quantum/classical routine is represented using a data structure. In Qiskit, the \verb|QuantumCircuit| class represents a program that can include both quantum and simple dynamic classical control.

\subsubsection{Hybrid Languages}

In a hybrid language, the programmer uses the same syntax for both classical or quantum operations.
Hybrid languages include CUDA-Q, Guppy, Q\#, and Silq. Q\# and Silq are standalone languages that provide their own compilers. CUDA-Q and Guppy are Python-based languages (also C++-based in the case of CUDA-Q), but they differ from the Python frameworks from Section~\ref{sec:embedded-languages}. In particular, they parse and compile annotated Python functions into their own intermediate representations, which enables static type checking and compiler optimizations.
All four hybrid languages use the same syntax for host control flow and dynamic control flow (i.e. \texttt{if} statements). Silq is the only language out of the four that also uses the same syntax for quantum control flow: \texttt{if} statements can also be used for quantum control.

Although we classify CUDA-Q and Guppy as hybrid, they share similarities with the embedded languages. In particular, they use Python as the classical host language. They allow the programmer to write interleaved quantum-classical functions in a hybrid style using their compilers, but they still require host-executed operations to be written in native Python. In contrast, Q\# and Silq make no distinction between host and dynamic operations---there is only one level of operation.

\begin{figure}[h]
\centering
\begin{lstlisting}[style=minted,language=Python]
@cudaq.kernel
def lcu_circuit():
    q = cudaq.qvector(2)
    success = False
    while not success:          # [All zero?]
        reset(q)                # [Reset]

        h(q[0])                 # [Prepare]
        cx(~q[0], q[1])         # [Select unitary (X)]
        cz(q[0], q[1])          # [Select unitary (Z)]
        h(q[0])                 # [Unprepare]

        success = not mz(q[0])  # [Measure]
\end{lstlisting}
\Description{}
\caption{In CUDA-Q, we can mix classical and quantum logic. In this example, we include the same repeat-until-success logic as in Figure~\ref{fig:example-qiskit-embedded-simple-1}, but we can use a typical Python while loop conditioned on a normal boolean expression. The comments with square brackets correspond to nodes in the LCU description in Figure~\ref{fig:algo-lcu}.}
\label{fig:example-cudaq-hybrid}
\end{figure}

CUDA-Q has a hybrid programming model, allowing the programmer to define quantum routines (called kernels) by writing normal Python functions annotated with the \verb|@cudaq.kernel| decorator. Inside of a kernel, the programmer may use a subset of Python, which is subsequently verified and compiled before execution on a quantum computer or simulator. Figure~\ref{fig:example-cudaq-hybrid} shows our CUDA-Q implementation of the repeat-until-success procedure in LCU. In contrast with the \verb|qc.while_loop()| method in Qiskit (Figure~\ref{fig:example-qiskit-embedded-simple-1}), we use a typical classical while loop. The loop condition can be any normal Python boolean expression, and the loop body can contain either classical or quantum operations. In CUDA-Q, there is little separation between quantum and classical operations. They can be seamlessly interleaved without much effort from the programmer, leaving the compiler to determine when and where to execute each operation.

\begin{figure}[t!]
\centering
\begin{lstlisting}[style=minted,language=Python]
sites = cirq.LineQubit.range(num_sites)
zz_terms = cirq.PauliSum()
x_terms = cirq.PauliSum()
for i in range(num_sites - 1):
    zz_terms += J * cirq.Z(sites[i]) * cirq.Z(sites[i+1])
for i in range(num_sites):
    x_terms += h * cirq.X(sites[i])

circuit = cirq.Circuit()
dt = time / steps
exp_zz = cirq.PauliSumExponential(zz_terms, exponent=-dt)
exp_x = cirq.PauliSumExponential(x_terms, exponent=-dt)
for _ in range(steps):
    circuit.append(exp_zz)
    circuit.append(exp_x)
\end{lstlisting}
\Description{}
\caption{Cirq enables easy construction of a Hamiltonian through the \texttt{PauliString} and \texttt{PauliSum} types as well as exponentiation of each term in the Hamiltonian.}
\label{fig:example-cirq-pauli}
\end{figure}

\subsection{Expressiveness}

Languages with high expressiveness make it easy for the programmer to implement algorithms at a high level. Our framework characterizes the expressiveness of a language using four different language features: support for Pauli manipulation, encoding of classical data, quantum integers, and dynamic allocation of quantum data.

Because many of the languages we discuss are provided as Python libraries, the line between native language features and library features is blurred. Therefore, we define a language feature as the union of the two---if a language natively supports a feature or provides the feature through its standard libraries, then we say that the language has that feature.

\subsubsection{Pauli Manipulation}

There are three types of Pauli manipulation we consider. First, the language provides data types that correspond to Pauli operators, Pauli strings, and sums of Pauli strings (i.e. Pauli sums). This allows the user to declaratively define a Hamiltonian, which is the first step in both of the Hamiltonian simulation algorithms shown in Figure~\ref{fig:algo-trotter} and Figure~\ref{fig:algo-lcu}. Second, the language supports exponentiation of Pauli strings or Pauli sums. This allows the user to easily program the classical exponentiation step of Trotterization from Figure~\ref{fig:algo-trotter}. Third, the language supports algebra on Hamiltonian operators, including addition, scalar multiplication, and operator-operator multiplication. This allows the user to easily compute the truncated Taylor expansion for use in the LCU algorithm, which is one of the classical preprocessing steps in Figure~\ref{fig:algo-lcu}.

Native Pauli manipulation is supported by Cirq, CUDA-Q, PennyLane, PyQuil, Qiskit, and Qrisp. These languages can be used to implement Hamiltonian simulation more declaratively using Pauli types, they alleviate the burden of exponentiating Pauli strings from the programmer, and they save many lines of code by computing the exponentiated Hamiltonian's Taylor expansion automatically.
In languages that do not support Pauli exponentiation, we manually implement each Hamiltonian's exponentiated terms; this requires few lines of code in the Trotterization benchmarks, but is significantly less flexible. In languages that do not support Pauli algebra, we manually compute the algebra using primitive data types; this requires many lines of code in the LCU benchmarks.

Cirq supports Pauli manipulation through its \verb|PauliString| and \verb|PauliSum| types. Figure~\ref{fig:example-cirq-pauli} shows how we perform Trotterization on the TFIM Hamiltonian in Cirq, encompassing all three parts of Figure~\ref{fig:algo-trotter}. First, we define the Hamiltonian. We start by initializing Pauli sums corresponding to each of the two sums in $H_\mathrm{TFIM}$. We append the individual terms of the sum, which correspond to Pauli $ZZ$ and $X$ operators applying to various sites. In Cirq, a single-qubit Pauli operator can defined on a specific site, and a Pauli string can be constructed by taking the tensor product of multiple Pauli operators using the \verb|*| operation. We can specify the real-valued coefficient of each Pauli string by simply multiplying it with a Python \verb|float| value.
Then we use Cirq's PauliSumExponential function to take the exponent of each sum in the Hamiltonian. Finally, we repeatedly apply them to the circuit for some number of steps.

\begin{figure}[t!]
\centering
\begin{lstlisting}[style=minted,language=Python]
zz_terms = PauliSum([])
x_terms = PauliSum([])
for i in range(num_sites - 1):
    zz_terms += J * (PauliTerm("Z", i) * PauliTerm("Z", i + 1))
for i in range(num_sites):
    x_terms += h * PauliTerm("X", i)

prog = Program()
dt = total_time / steps
exp_zz = exponentiate_commuting_pauli_sum(zz_terms)(dt)
exp_x = exponentiate_commuting_pauli_sum(x_terms)(dt)
for _ in range(steps):
    prog += exp_zz
    prog += exp_x
\end{lstlisting}
\Description{}
\caption{PyQuil is similar to Cirq in that it allows the user to easily construct and exponentiate a Hamiltonian consisting of a sum of Pauli strings.}
\label{fig:example-pyquil-pauli}
\end{figure}

In PyQuil, Trotterization is programmed in the same way as Cirq (Figure~\ref{fig:example-pyquil-pauli}). One minor difference is that the sites are indexed using integers, rather than qubit types. Here, each single-qubit Pauli operator is represented by the \verb|PauliTerm| type, and we can define interaction terms (such as the ZZ interaction) by multiplying them together.
Like Cirq, we can add a coefficient to each Pauli string by multiplying it with a Python \verb|float| value.

\begin{figure}[h]
\centering
\begin{lstlisting}[style=minted,language=Python]
H = 0
for i in range(0, n_sites-1):
    H += coupling * spin.x(i) * spin.x(i + 1)
    H += coupling * spin.y(i) * spin.y(i + 1)
    H += coupling * spin.z(i) * spin.z(i + 1)
for i in range(0, n_sites):
    H += field * spin.z(i)
taylor_H = spin.i(0) - (1j * H * t) - (H * H * t**2 / 2)
\end{lstlisting}
\Description{}
\caption{CUDA-Q allows the programmer to not only construct a Hamiltonian from Pauli operators, but also perform algebra on the Hamiltonian. This lets us calculate the Taylor series expansion for LCU in only a single line of code.}
\label{fig:example-cudaq-pauli}
\end{figure}

Figure~\ref{fig:example-cudaq-pauli} shows a key part of the LCU algorithm in CUDA-Q, implementing the ``Classical Preprocessing'' part of Figure~\ref{fig:algo-lcu}. In this example, we construct the Heisenberg Hamiltonian $H_\mathrm{Heis}$ using CUDA-Q's provided \verb|spin| operators. Since spin operators are equivalent to Pauli operators up to a scaling factor, we henceforth refer to the operators in this example as Pauli operators.
The X, Y, Z, and identity Pauli operators are defined as objects in CUDA-Q's spin module. This allows for type checking of these operators; this is contrasted with Figure~\ref{fig:example-cirq-pauli} and Figure~\ref{fig:example-pyquil-pauli}, in which the programmer specifies the Pauli operators using strings.
CUDA-Q also supports algebra on a Pauli operator. In the last line of the example, we compute the first two non-constant terms of the Taylor expansion of $e^{-iHt}$. The \verb|spin| operator data type allows us to add them together, multiply them together, and multiply them with scalar coefficients. In a language without support for this kind of operator algebra, squaring a Hamiltonian requires representing it as a sum of Pauli strings and multiplying each of the possible $n^2$ pairs of Pauli strings. In practice, being able to perform this operation in a single line of code saves a significant amount of programmer effort.

\subsubsection{Encoding Classical Data}

There are two main ways to encode classical data in a quantum circuit. A nonzero complex-valued vector with length less than $2^n$ can be encoded as the state vector of $n$ qubits, up to a constant factor, using a method called \textit{state preparation}. This requires the initial $n$-qubit state to be known and unentangled with any other qubits. Similarly, any $2^n\times 2^n$ complex-valued square matrix can be embedded in the top-left corner of the unitary matrix representation of a quantum circuit using a method called \textit{block encoding}. A common way to implement this involves representing the matrix as a linear combination of unitary matrices, encoding the coefficients using state preparation, and encoding the unitary matrices as quantum operations.

Vector or matrix encoding is supported by Cirq, PennyLane, Q\#, Qiskit, Qualtran, and Qrisp. CUDA-Q supports initializing the statevector of a set of qubits, but does not synthesize the state preparation as a quantum circuit. Guppy does not natively support state preparation, but allows the user to import a state preparation operation from pytket. Silq has a separate repository that contains implementations of state preparation and block encoding, but these are not provided as a standard library. All of the above languages, with the three caveats, support encoding a vector as a quantum state. In addition, Qualtran and Qrisp support block-encoding a matrix as a quantum circuit. In languages that don't support vector encoding, state preparation must be implemented manually through recursive qubit rotations. In languages that don't support encoding a matrix, block encoding must be implemented manually through a prepare-select-unprepare routine along with repeat-until-success, as shown in Figure~\ref{fig:algo-lcu}.

\begin{figure}[h]
\centering
\begin{lstlisting}[style=minted,language=Python]
qp.MottonenStatePreparation(amps, wires=indices)
\end{lstlisting}
\Description{}
\caption{PennyLane provides a library function to perform state preparation on a set of qubits, which are indexed with integers.}
\label{fig:example-pennylane-encode}
\end{figure}

As shown in Figure~\ref{fig:example-pennylane-encode}, PennyLane provides an operation that allows the user to easily encode a vector of complex numbers (called \verb|amps| in this example) into the state vector of some qubits. Here, the qubits are specified as a list of integer indices, where each index corresponds to one qubit. This performs the state preparation algorithm introduced by M{\"o}tt{\"o}nen et al. in Ref.~\cite{mottonen2004transformation}. PennyLane provides the \verb|MottonenStatePreparation| function directly from the top-level module called \verb|qp|, from which all basic functions and classes can be imported. This function is used in the ``Prepare'' operation in LCU shown in Figure~\ref{fig:algo-lcu}.


\begin{figure}[h]
\centering
\begin{lstlisting}[style=minted,language=qsharp]
open Std.StatePreparation;
operation ApplyLCUBlock(...) {
    ...
    PreparePureStateD(amplitudes, qubits);
    ...
}
\end{lstlisting}
\Description{}
\caption{Q\# provides a library function to perform state preparation on a set of qubits, which are passed in as a native qubit type.}
\label{fig:example-qsharp-encode}
\end{figure}

Figure~\ref{fig:example-qsharp-encode} shows how Q\# provides state preparation as a library function. The programmer must import the \verb|Std.StatePreparation| module, which allows them to use the \verb|PreparePureStateD| function. This function is used in mostly the same way as PennyLane's state preparation function (Figure~\ref{fig:example-pennylane-encode}). The only difference is that Q\# provides a qubit type, which allows the programmer to pass in a list of qubits into the function, rather than a list of qubit indices as in PennyLane. This also implements ``Prepare'' in Figure~\ref{fig:algo-lcu}.


\begin{figure}[h]
\centering
\small
\begin{lstlisting}[style=minted,language=Python]
def build_lcu_block(
    paulis: List[cirq.DensePauliString],
    weights: List[float],
    precision: float,
) -> LCUBlockEncoding:
    target_bitsize = len(paulis[0])
    selection_bitsize = ceil(log2(len(paulis)))
    
    select = TaylorSelectPauliLCU(selection_bitsize, target_bitsize, paulis)
    prepare = StatePreparationAliasSampling.from_probabilities(weights, precision)
    
    return LCUBlockEncoding(
        select=BlackBoxSelect(select), prepare=BlackBoxPrepare(prepare)
    )
\end{lstlisting}
\Description{}
\caption{Qualtran provides an LCUBlockEncoding operation. When combined with the BlackBoxSelect and BlackBoxPrepare operations, this enables the user to block-encode a linear combination of any unitary operations.}
\label{fig:example-qualtran-encode}
\end{figure}

Qualtran provides operations to easily block-encode a matrix using the LCU algorithm. Figure~\ref{fig:example-qualtran-encode} is an example of how we use Qualtran's \verb|LCUBlockEncoding| for Hamiltonian simulation. This implements the ``Quantum Circuit'' part of Figure~\ref{fig:algo-lcu}. After we have computed the Taylor series expansion of $e^{-iHt}$ and represented it as a weighted sum of Pauli strings, we want to block-encode the weighted sum into the top-left corner of the unitary matrix of a quantum operation. First, we must convert the Pauli strings into selected quantum operations using the custom \verb|TaylorSelectPauliLCU| block. Qualtran does provide a similar block called \verb|SelectPauliLCU|, but this only supports real coefficients; we need to preserve the full complex phase of each Pauli string. Next, we use Qualtran's \verb|StatePreparationAliasSampling| to encode the weights of our Pauli strings. By using the \verb|from_probabilities| method, we can provide the weight vector as-is. This is different from the amplitude vectors from Figure~\ref{fig:example-pennylane-encode} and Figure~\ref{fig:example-qsharp-encode}, which are the square roots of the desired weights. Finally, Qualtran enables us to easily create a block encoding operation using LCU with our selection and preparation operations, which correspond to the unitaries and their coefficients, respectively.

\subsubsection{Quantum Integers}

In languages that support quantum integers, they are typically represented as a special kind of integer type that can be manipulated using the typical arithmetic operators. An $n$-qubit quantum integer is represented as $n$ qubits, and arithmetic operations on them act on each of the $2^n$ computation basis states simultaneously. When a program using quantum arithmetic is synthesized into a circuit, the compiler will convert operations into quantum gates using techniques like quantum adders. This saves a significant amount of manual programming for algorithms that require quantum arithmetic, such as Shor's algorithm.

Quantum integers are supported by Qualtran, Qrisp, and Silq. In languages that don't support quantum integers, the programmer must implement modular exponentiation by using untyped quantum modular arithmetic libraries, decomposing a permutation matrix (which does not scale), or implementing quantum arithmetic themselves. We use modular arithmetic libraries for Cirq, PennyLane, and Q\#. We use permutation matrices for CUDA-Q, Guppy, PyQuil, and Qiskit because these languages do not support quantum modular arithmetic in their standard libraries.

\begin{figure}[h]
\centering
\begin{lstlisting}[style=minted,language=Python]
def find_order(a, N, t):
    qg = QuantumModulus(N)
    qg[:] = 1
    qpe_res = QuantumFloat(t, exponent=-(t))
    h(qpe_res)
    for i in range(len(qpe_res)):
        with control(qpe_res[i]):
            qg *= a
            a = (a * a) % N
    QFT(qpe_res, inv=True)
    return qpe_res, qg
\end{lstlisting}
\Description{}
\caption{Qrisp supports quantum floats. It also has a QuantumModulus type, which allows the user to easily perform modular arithmetic on quantum variables. This is very useful in Shor's algorithm, which requires quantum modular exponentiation.}
\label{fig:example-qrisp-qint}
\end{figure}

Qrisp has two data types that are useful for quantum arithmetic, which are both shown in the example in Figure~\ref{fig:example-qrisp-qint}. \verb|QuantumFloat| is used to represent a quantum floating point number. The user can specify the qubit width of the significand, as well as the exponent. These can be used as quantum integers by setting the exponent to zero. \verb|QuantumModulus| is used to represent a quantum integer under modular arithmetic. The user specifies the modulus $N$, and the result of every operation on the qubits will be modulo $N$. Notice that the user does not need to specify the number of qubits that are used to store the number. This is because Qrisp determines the required number of qubits to prevent overflow based on $N$ automatically. In this example, the modular exponentiation routine of Shor's algorithm is implemented using \verb|QuantumModulus|. This example also represents the $t$ counting qubits as a \verb|QuantumFloat| with exponent $-t$ because the algorithm interprets them as a real number in the interval $[0,1)$. In relation to the diagram in Figure~\ref{fig:algo-shors}, this example implements the ``Quantum Circuit'' part. \verb|qpe_res| are the counting qubits, and they are prepared with \verb|h(qpe_res)|. The loop implements the modular exponentiation procedure, and \verb|QFT(..., inv=True)| implements the inverse Quantum Fourier Transform.

\begin{figure}[t!]
\centering
\begin{lstlisting}[style=minted,language=silq]
// a^b mod num
def powm[n:!N,m:!N](a:uint[n], b:uint[m], num:!N)lifted: uint[n] {
    (r, x) := (1:uint[n], a);

    for i in [0..m) {
        if b[i] {
            r = r*x % num;
        }
        x = x*x % num;
    }

    return r;
}
\end{lstlisting}
\Description{}
\caption{Silq supports quantum integers, which allows the user to easily program the modular exponentiation routine in Shor's algorithm. Silq supports quantum versions of many of its classical data types, and arithmetic operations can be used between both of the types transparently.}
\label{fig:example-silq-qint}
\end{figure}

In Silq, many of the classical data types also have quantum equivalents. Figure~\ref{fig:example-silq-qint} shows Shor's algorithm's modular exponentiation routine implemented with repeated quantum multiplication. This implements the ``Modular exponentiation'' part of Figure~\ref{fig:algo-shors}. Silq's \verb|uint[n]| data type represents an $n$-qubit quantum integer, whereas \verb|B^n| represents a vector of $n$ qubits. In principle, these two data types are represented in the same way in hardware. The difference is in which operations are supported; for example, quantum integers support arithmetic, and quantum vectors support concatenation and truncation. In Silq, the classical versions of data types are denoted with an exclamation point (i.e. \verb|!uint[n]| is a classical $n$-bit integer).

\subsubsection{Dynamic Allocation of Qubits}

Languages that support dynamic allocation of qubits typically expose such behavior through a \verb|qubit_init()| function. An important concern with dynamic allocation of data is ensuring that data is freed once it is no longer being used. A language can support this by requiring the programmer to call a \verb|qubit_free()| function, by providing scopes that automatically allocate and free temporary qubits, or by enforcing ownership semantics that prevent the programmer from accidentally leaking quantum data. 
We say that a language supports dynamic allocation if qubits can be allocated at run time, depending on mid-circuit measurements.

Dynamic qubit allocation is fully supported in Guppy and Silq. It is partially supported in CUDA-Q, PennyLane, Q\#, Qrisp; these languages allow the programmer to allocate qubits in the middle of the program, but allocation may not depend on quantum routine's behavior.
In languages that do not support dynamic allocation, the programmer must know upfront exactly how many qubits they will need and must allocate them manually. Dynamic allocation is not strictly necessary for our examples, but it simplifies the programs by allowing us to create temporary quantum variables when we need them and discard them after we are done using them.

\begin{figure}[t!]
\centering
%
%
%
%
\begin{lstlisting}[style=minted,language=Python]
@guppy()
def lcu() -> None:
    opr = array(qubit() for _ in range(comptime(n)))

    while True:
        ctrl = array(qubit() for _ in range(comptime(m)))
        for i in range(len(opr)):
            reset(opr[i])
        # Omitted: [Prepare], [Select Unitaries], [Unprepare]
        
        measurement = measure_array(ctrl)
        # Omitted: if [All zero?] then break

    discard_array(opr)
\end{lstlisting}
\Description{}
\caption{Guppy supports dynamic allocation of qubits, which is useful in this LCU repeat-until-success loop. At the beginning of each loop, we allocate some ancilla qubits, and at the end of the loop, we measure them and conditionally perform another loop iteration.}
\label{fig:example-guppy-dyn}
\end{figure}

Guppy allows the user to dynamically allocate and free qubits. Figure~\ref{fig:example-guppy-dyn} shows how we implement the repeat-until-success procedure used in the LCU algorithm with some omissions labeled according to Figure~\ref{fig:algo-lcu}. This example shows four operations related to qubit allocation that Guppy provides: allocation, reset, measurement, and discarding. Allocation is performed with the \verb|qubit()| function, which can be performed dynamically depending on the runtime behavior of the quantum program. In this example, the \verb|while| loop's break condition is dynamic, so the allocation of \verb|ctrl| is dynamic. Resetting a qubit to $\ket{0}$ is also supported via the \verb|reset()| function. \verb|measure()| and \verb|discard()| are the two qubit deallocation functions provided by Guppy. Guppy's type checker will ensure that every qubit the programmer allocates is either measured or discarded before the end of the program. In this example, we use all four operations. For the \verb|opr| qubits, we allocate them once and reset them at the beginning of each loop. For the \verb|ctrl| qubits, we cannot do the same because at the end of the loop, they must be measured to determine whether to break out of the loop. This means they must be dynamically allocated at the beginning of each loop iteration. At the end of the function, we must discard \verb|opr| to ensure the quantum resources are not leaked.

\begin{figure}[h]
\centering
%
%
%
%
%
\begin{lstlisting}[style=minted,language=silq]
while true {
    qs := vector(n, 0:B);
    ancilla := vector(n_ancilla, 0:B);
    // Omitted: [Prepare], [Select], [Unprepare]

    measurement = measure(ancilla);
    // Omitted: if [All zero?] then break
    
    measure(qs);
}
\end{lstlisting}
\Description{}
\caption{Silq allows dynamic allocation of qubits. Vectors of quantum types can be initialized with the \texttt{vector} constructor, and quantum data can be deallocated via measurement.}
\label{fig:example-silq-dyn}
\end{figure}

In Silq, quantum and classical data can be allocated by simply initializing a variable. For initializing dynamically-sized arrays or fixed-size vectors of qubits, Silq provides the \verb|array| and \verb|vector| constructors. Figure~\ref{fig:example-silq-dyn} shows an implementation of the repeat-until-success loop in the LCU algorithm, similar to the Guppy implementation in Figure~\ref{fig:example-guppy-dyn}. The main difference is that \verb|qs| (called \verb|opr| in Figure~\ref{fig:example-guppy-dyn}) is allocated and deallocated at the beginning and end of each loop, rather than declaring the variable outside of the loop and resetting it at the beginning of each loop. In the loop, we also initialize a vector of ancilla qubits, and at the end of the loop, we measure the ancilla qubits, discarding them. Finally, we either break from the loop or perform another iteration. If we break, then we can return the modified \verb|qs| qubits. If we perform another iteration, then we must first discard the existing \verb|qs| qubits by measuring them. Like Guppy, Silq ensures that data is not leaked by requiring the programmer to either discard quantum variables explicitly or return them. A key difference, however, is that Guppy requires explicit deallocation before the end of a function through the use of \verb|discard|, whereas Silq can usually automatically discard and cleanup resources automatically. The exact situations when Silq can do this are discussed in Section~\ref{sec:initialization-safety}.

\subsection{Safety}

Languages with high safety prevent the programmer from making mistakes that result in classical or quantum data in a bad state. We characterize the safety of quantum programming languages with two dimensions: static type checking and initialization safety.

\subsubsection{Static Type Checking}

Static type checking is used to verify before the program executes that the program is well-typed. Type checking in quantum programming languages is fundamentally the same as in classical programming languages, except with the introduction of quantum types and operations that have their own typing rules. Many of the languages we discuss are based on Python, which does not natively support static type checking, but some of the Python-based languages provide their own static type checking implementations. This is because these languages construct a quantum program by compiling a quantum program into a circuit, and then executing the circuit on a quantum computer or simulator. During this compilation process, the language may perform type checking before the quantum program begins execution.
Python does natively support type annotations, but these must be checked with an external static type checking tool, so we do not count this as language-level support for static type checking.

Static type checking is supported by the standalone languages Q\# and Silq. CUDA-Q also naturally supports it when C++ is used as the host language. Among the Python-based languages, CUDA-Q and Guppy type check and compile quantum programs before execution.
In addition, Guppy and Silq enforce linear type checking, which provides some guarantees that programs are physically realizable and that quantum resources can be managed safely.

\begin{figure}[h]
\centering
\begin{lstlisting}[style=minted,language=Python,basicstyle=\ttfamily\small]
n = guppy.nat_var("n")  # Generics let us do parametrization

@guppy
def tfim_trotter_step(qs: array[qubit, n], J: float, h: float, dt: float) -> None:
    zz_angle = angle(2 * J * dt / comptime(math.pi))
    for i in range(n-1):
        evolve_zz(qs[i], qs[i+1], zz_angle)
        
    x_angle = angle(2 * h * dt / comptime(math.pi))
    for i in range(n):
        rx(qs[i], x_angle)

tfim_trotter_step.check()  # Do type checking before the function executes
\end{lstlisting}
\Description{}
\caption{Guppy type checks functions with the \texttt{@guppy} decorator before they are executed. In this example, we must use Python type annotations to specify the types of the function arguments as well as the return type. Array types have static element types and sizes, so array type annotations must be parametrized with both. Here, the \texttt{qs} function parameter is an array type, where each element is a qubit, and the array length is a generic parameter \texttt{n}. Guppy also lets us invoke type checking manually before compiling the function with the \texttt{check()} method.}
\label{fig:example-guppy-type}
\end{figure}

In Guppy, a quantum function is annotated with the \verb|@guppy| decorator. This tells Guppy that the function may only be called from another quantum function or in the context of a quantum processor or simulator. Figure~\ref{fig:example-guppy-type} shows an example of a quantum subroutine that is called from another quantum function, and it implements a part of Trotterization.
In relation to Figure~\ref{fig:algo-trotter}, this code implements exponentiation of each Hamiltonian term by time \verb|dt|\,$=t/r$ as well as applying the exponentiated terms to some qubits; it does not include an explicit definition of the Hamiltonian nor the $r$-times repetition.
In this example, we define a function that takes a set of $n$ qubits and performs a single Trotterization step. We must use Python's type annotations to specify the types of the function parameters at compile time. The \verb|qs| parameter is an array of qubits, which is annotated as an \verb|array| type and must have a static element type and size. To support parametrization of the array size, Guppy allows us to create a generic size parameter using \verb|guppy.nat_var()|. Inside of the quantum function, most of the code is executed at runtime on the quantum executor, so Guppy only supports a subset of Python features. However, we can use \verb|comptime()| to evaluate an expression at compile time, which gives us the full Python execution environment.
Besides single expressions, Guppy allows function arguments, as well as entire functions themselves, to be annotated as compile-time.

\subsubsection{Initialization Safety}
\label{sec:initialization-safety}

Initialization safety ensures that all resources are initialized in a clean state. In classical computing, this means newly declared variables cannot be left uninitialized. Classical programming languages can support this by defining a default value for every primitive data type, and setting every new variable to the type's default value. In quantum computing this is not as simple because it is not always possible to set acquired resources to a clean state. Previously-discarded qubits may be entangled with other qubits that are currently in-use, so attempting to acquire these qubits and reset them to a known value could have non-deterministic effects on the state of the program. To ensure that quantum resources can be acquired and initialized safely, a quantum programming language must enforce that qubits are uncomputed before they are discarded. This can be done automatically under certain conditions, which can be determined by a compiler ahead of time.

Initialization Safety is supported by Qrisp and Silq. Qrisp provides the \verb|@auto_uncompute| function decorator, and Silq enforces initialization safety through its type system.

Figure~\ref{fig:example-silq-qint} shows a Silq function that performs modular exponentiation on quantum variables \verb|a| and \verb|b|; it also uses helper variables \verb|r| and \verb|x|. 
This implements the ``Modular exponentiation'' part of Figure~\ref{fig:algo-shors}.
The function is annotated as \verb|lifted|, which means its arguments are treated as read-only controls and it does not contain any quantum operations. Note that Silq defines a ``quantum operation'' to be one that introduces or destroys superpositions, and that Silq will recognize that the function in Figure~\ref{fig:example-silq-qint} contains only classical operations on quantum data. In Silq, any \verb|lifted| function or expression can be uncomputed, which means any temporary variables are automatically cleaned up. In this example, \verb|x| is a temporary variable because it is initialized in the function, and it is not returned. Therefore, at the end of the function, Silq automatically uncomputes \verb|x|, unentangling it from other values and allowing its resources to be cleanly discarded. If the compiler decides to reuse these qubits later in the program, it can acquire and initialize them without worrying about non-deterministic effects on other parts of the program.

\subsection{Summary}

In this section, we have presented examples of how certain language features can make languages easier to use, more expressive, or safer. Though every language has strengths and weaknesses, here we present a non-exhaustive list of highlights. CUDAQ's Pauli operator types make it very easy to exponentiate Pauli strings as well as perform algebra on them to compute the truncated Taylor series expansion. Qualtran provides a high-level and easy to use block encoding operation. Qrisp's \verb|QuantumModulo| type makes it easy to implement modular exponentiation for Shor's algorithm. Lastly, Silq enforces both type safety and initialization safety at compile time.


\section{Experimental Comparison}
\label{sec:experimental-comparison}

This section describes our experimental methodology and our results.
Our benchmark programs, testing harness, and configuration files are available in our public artifact repository~\cite{github_repo}. 
Our experiments were performed on macOS with an Apple Silicon M5 Pro chip and 24 GB of memory.

\begin{table*}[t]
  \centering
  \caption{How We Execute the Programs in the Ten Languages}
  \label{tab:how-we-execute-programs}
  \begin{tabular}{l|c|c|c}
    \toprule
             & Classical Host    & Intermediate      &   \\
    Language & Executor (if any) & representation (if any) & Simulator  \\
    \midrule
    Cirq              & Python & Cirq circuit & \href{https://google.com}{Cirq simulator} \\
    CUDA-Q            & Python & Quake, QIR~\cite{lubinski2022qir} & \href{https://nvidia.github.io/cuda-quantum/latest/using/backends/sims/svsims.html#cpu}{qpp-cpu statevector simulator} \\
    Guppy             & Python & HUGR \cite{Koch2025Hugr} & \href{https://quest.qtechtheory.org/}{QuEST simulator}~\cite{jones2019quest} \\
    PennyLane         & Python & PennyLane Quantum Tape & \href{https://pennylane.ai/devices/default-qubit}{\texttt{default.qubit} device} \\
    PyQuil            & Python & Quil program & \href{https://pyquil-docs.rigetti.com/en/stable/apidocs/pyquil.simulation.html}{NumpyWavefunction simulator} \\
    Q\#               & (compiled) & QIR~\cite{lubinski2022qir} & \href{https://learn.microsoft.com/en-us/azure/quantum/sparse-simulator}{QDK sparse simulator} \\
    Qiskit            & Python &  Qiskit QuantumCircuit & \href{https://qiskit.github.io/qiskit-aer/stubs/qiskit_aer.StatevectorSimulator.html}{Qiskit statevector simulator} \\
    Qualtran          & Python & Cirq circuit & \href{https://quantumai.google/reference/python/cirq/Simulator}{Cirq simulator} \\
    \hline
    Qrisp             & Python & Qrisp circuit & \href{https://github.com/eclipse-qrisp/Qrisp/blob/1aada76a958637eb448a769aaeb4da24fec98bd8/src/qrisp/simulator/simulator.py#L230}{Qrisp statevector simulator} \\
    Silq              & (interpreted) & (none) & \href{https://github.com/silq-lang/silq/blob/master/qsim.d}{Silq QSim simulator} \\
    \bottomrule
  \end{tabular}
\end{table*}

Table~\ref{tab:how-we-execute-programs} summarizes how we execute our programs in the ten languages, which in all cases is by simulation.
The table also lists the intermediate representations we use for each language. This is not exhaustive---for example, Qrisp supports \href{https://www.qrisp.eu/reference/Jasp/index.html}{MLIR} and Silq supports \href{https://github.com/silq-lang/silq/blob/master/hqir.d}{HQIR}, but we do not use these in our implementations.

Table~\ref{tab:experimental-results} reports the execution times, averaged over twenty runs. We measured the full execution process of each program, including library imports, quantum circuit compilation, and simulation. These execution times vary across languages but are all low enough to support a productive programming experience. 
We enable a fair and deterministic comparison of the execution times by reporting the time to execute only one run of the quantum routine in Shor's algorithm and a single iteration of the repeat-until-success loop in LCU.
We emphasize that Table~\ref{tab:experimental-results} is not a definitive performance ranking. It measures the execution times of small experiments, including various overheads that would disappear in real-world programs. The purpose of the table is to show that all languages can actually be executed to run experiments with reasonable compilation and execution times.

Table~\ref{tab:experimental-results} lists several execution times that seem long, for reasons that we discuss next.
Guppy's long execution times for Shor's algorithm and LCU are a result of our programs importing pytket operations and converting them into Guppy function definitions, as well as our inclusion of type checking and compilation times. PennyLane's long execution time for Shor's algorithm is a result of how PennyLane implements quantum modular arithmetic---for this small benchmark, it results in a slower quantum circuit. Qiskit's long execution time for Shor's algorithm compared to the Hamiltonian simulation benchmarks is due to our permutation matrix implementation, which spends most of the time decomposing a permutation matrix into a controlled unitary oracle. Qualtran's long Shor's algorithm execution time is due to how it decomposes the \texttt{ModExp} operation---the resulting Cirq circuit is slower than our Cirq implementation, which uses \texttt{cirq.ArithmeticGate}. Qualtran's long LCU execution times are because of our use of state preparation via alias sampling, rather than via controlled rotation gates; we chose this method because of its compatibility with Qualtran's LCU operation.

All the programs produce correct results.
For Shor's algorithm, we factor $N = 21$ using five operand qubits and six ancilla qubits, and we check that the output is a correct factor. Because Shor's algorithm can succeed purely by chance, we also verify that the quantum state of each program matches that of a reference implementation.
For Hamiltonian simulation, we simulate four three-qubit Hamiltonians per language and check that the quantum state has fidelity at least 0.99 compared to a reference result computed using NumPy.

\begin{table*}[t]
  \centering
  \caption{Execution Times in Seconds}
  \begin{tabular}{c|r|rr|rr}
    \toprule
             & \multicolumn{1}{c|}{Factoring}
             & \multicolumn{2}{c|}{TFIM} & \multicolumn{2}{c}{Heisenberg} \\
    Language & Shor's alg. & Trotter & LCU & Trotter & LCU \\
    \midrule
    Cirq      &  0.6\,\, & 0.6 & 0.6 & 0.6 &  0.6 \\
    CUDA-Q    &  0.9$^1$ & 0.7 & 0.8 & 0.7 &  0.8 \\
    Guppy     & 11.9$^1$ & 1.3 & 2.5 & 0.9 & 11.7 \\
    PennyLane &  9.9\,\, & 1.9 & 1.7 & 1.7 &  1.7 \\
    PyQuil    &  0.2$^1$ & 0.2 & 0.2 & 0.1 &  0.3 \\
    Q\#       &  0.1\,\, & 0.1 & 0.1 & 0.1 &  0.1 \\
    Qiskit    &  1.8$^1$ & 0.1 & 0.1 & 0.1 &  0.3 \\
    Qualtran  & 13.6\,\, & 1.0 & 2.7 & 1.0 & 11.8 \\
    \hline
    Qrisp     &  2.9\,\, & 1.4 & 1.1 & 1.0 &  1.1 \\
    Silq      &  0.2\,\, & 0.1 & 0.1 & 0.1 &  0.1 \\
    \bottomrule
  \end{tabular}
  \label{tab:experimental-results}
  \caption*{
    $^1\,$These implementations use a permutation matrix, rather than quantum modular arithmetic.
  }
\end{table*}

\section{Open Challenges and Future Directions}
\label{sec:open-challenges-and-future-directions}

\paragraph{Challenges for Future Language Design.}
Our conceptual comparison of the ten languages highlighted that no single language supports all the expressiveness and safety characteristics.  For example, among the embedded languages, Qrisp supports most of the characteristics but is missing type safety and full dynamic qubit allocation.  Similarly, among the hybrid languages, Silq supports most of the characteristics but is missing native support for Pauli manipulation and for encoding of classical data.
For both kinds of languages, an open challenge is to either enhance an existing language or design a new language that has all of the characteristics.
Another challenge is to support emerging language ideas, such as assertions
\cite{10.1145/3428218},
program verification
\cite{10.1145/3453483.3454061}, and
distributed quantum programming
\cite{10.1145/3649329.3655908}.
Those ideas have been studied in a variety of contexts, including some of the languages in our survey, and they are interesting directions for future language design.

\paragraph{Adding Languages and Benchmarks.}
Researchers can easily apply our methodology to additional languages and benchmarks.  They will find our classification framework easy to use and they will find our GitHub repository easy to extend.  
In particular, they can easily extend our scripts to run all the tests for a new language and run a new test for all the languages.
The result is that when future languages and benchmarks emerge, researchers can do a low-effort comparison with the state of the art.

\paragraph{Survey of Language Implementations.}
Our survey focuses on language design but ignores language implementation beyond simulators.  Throughout our survey effort, we noticed the wide variety of implementations that make quantum programming languages run on quantum computers.  For example, those implementations differ in their support for fault tolerance, in their approach to optimization, and in their capabilities for targeting multiple kinds of quantum computers. We leave to future work to survey those implementations.



\section{Conclusion}
\label{sec:conclusion}

We presented a survey of quantum programming languages through the lens
of five important workloads. Specifically, we compared ten high-level, general-purpose, executable languages that are available as actively maintained open-source software, and we found that no single language has every desirable characteristic.  We arrived at this conclusion via use of a classification framework, a suite of programs, and a testing harness, all of which can be useful for evaluating future language designs.  Our classification framework has dimensions that arose from the needs of Shor's algorithm and Hamiltonian simulation. Future work may introduce new dimensions, perhaps based on the needs of other important quantum algorithms.

\paragraph{Acknowledgments.}
We are grateful to Christopher Cherfan, Timon Gehr, Olivia Di Matteo, Raphael Seidel, Kartik Singhal, Martin Vechev, Hristo Venev, and Charles Yuan for many helpful comments on drafts of our paper and GitHub directory.
We are supported by the NSF QLCI program through grant number OMA-2016245 and the NSF NQVL program through grant number 2435382.


\bibliographystyle{ACM-Reference-Format}
\bibliography{references}

\end{document}